\documentclass[aip,apl,reprint]{revtex4-1}
\usepackage{amsmath,graphicx}
\usepackage{color,ulem}

\definecolor{darkred}{rgb}{0.90,0,0}
\definecolor{darkgreen}{rgb}{0,0.60,.2}
\definecolor{darkblue}{rgb}{0,0,1}
\definecolor{grey}{cmyk}{0,0,0,0.25}
\definecolor{orange}{cmyk}{0,0.6,0.8,0}

\newcommand{\be}{\begin{equation}}
\newcommand{\ee}{\end{equation}}

\begin{document}

\title{Towards spin injection from silicon into topological insulators: Schottky barrier between Si and Bi$_2$Se$_3$}
\author{C. Ojeda-Aristizabal }
\author{M. S. Fuhrer }
\affiliation{Center for Nanophysics and Advanced Materials,
University of Maryland, College Park, MD 20742-4111, USA}
\author{N. P. Butch}
\altaffiliation{Condensed Matter and Materials division, Lawrence Livermore National Laboratory, Livermore, CA 94550, USA (present address)}
\author{J. Paglione}
\author{I. Appelbaum}
\affiliation{Center for Nanophysics and Advanced Materials,
University of Maryland, College Park, MD 20742-4111, USA}

\begin{abstract}
A scheme is proposed to electrically measure the spin-momentum coupling in the topological insulator surface state by injection of spin polarized electrons from silicon. As a first approach, devices were fabricated consisting of thin ($<$100nm) exfoliated crystals of Bi$_2$Se$_3$ on n-type silicon with independent electrical contacts to silicon and Bi$_2$Se$_3$. Analysis of the temperature dependence of thermionic emission in reverse bias indicates a barrier height of 0.34 eV at the Si-Bi$_2$Se$_3$ interface. This robust Schottky barrier opens the possibility of novel device designs based on sub-band gap internal photoemission from Bi$_2$Se$_3$ into Si.

\end{abstract}

\maketitle

The most remarkable feature of the three-dimensional strong topological insulators (TIs) is the existence of a metallic surface state within the bulk bandgap with chiral charge carriers exhibiting perfect spin-momentum coupling. The chiral TI surface state has been proposed as the basis of spintronics and quantum computing devices. \cite{Xia_NatPhys2009, Hasan_Rev2012}

  Although the spin helicity of the TI surface has been experimentally measured by spin-angle resolved photoemission spectroscopy (spin-ARPES) \cite{Pan_PRL2011, Hsieh_Nature2009,Wang_PRL2011}, no electrical transport experiment has yet shown clear evidence of it for several reasons. Since spin and momentum are perfectly coupled, non local measurement of a spin current in the absence of a charge current \cite{Lou_NatPhys2007, Sasaki_APL2010} is precluded. Perhaps even more importantly, spin precession induced by a weak perpendicular magnetic field is also eliminated by momentum scattering in the diffusive regime. This is particularly problematic since evidence of spin precession and dephasing are used to unambiguously identify spin transport in both inorganic semiconductors \cite{Appelbaum_Nature2007} and metals \cite{Johnson_PRL1985}.

Direct measurement of the spin Hall effect by injection of spin from a ferromagnetic (FM) contact is also problematic since the desired signal will be difficult to distinguish from the ordinary Hall effect due to stray fields from the nearby FM electrode.

  More sophisticated measurement geometries are unlikely to solve the problem. For instance, if the FM magnetization is oriented by an in-plane magnetic field at an angle with the charge current, anisotropic magnetoresistance results. This gives rise to a planar Hall effect \cite{Goldberg_PR1954, Bullis_PR1956,CampbellFert_1982} which is difficult to distinguish from a signal due to the spin-momentum coupling in the TI \cite{Xia_Arxiv2012}.  The situation would not be different if the current flows perpendicular to the interface between TI and the ferromagnet, since the strong spin orbit interaction of the TI may induce an anisotropic tunneling magnetoresistance signal, as has been observed in devices consisting of a tunneling barrier between a ferromagnet and a non-magnetic layer. \cite{Moser_PRL2007, Uemura_APL2011}

Here, we propose a class of transport experiments to confirm the spin-momentum coupling in TI surface states which circumvents these problems by injecting spin polarized electrons from a long-distance silicon transport channel into the topological insulator. Detection of the spin-momentum coupling is provided by the intrinsic spin Hall effect in the TI surface in the ballistic regime. The significant advantages of this scheme are (1) the separation of the FM spin source from the TI, made possible by the long spin transport distances achievable in undoped silicon, eliminating stray field effects and magnetoresistance in the TI contacting the silicon; and (2) the possibility of magnetic field-controlled spin precession in the Si channel which allows spin direction control without contact magnetization reversal. Thus the intermediate transport channel provides an unambiguous means to confirm the spin-dependence (and magnetization-independence) of the resulting spin-Hall signals in the TI.

\begin{figure}
    \includegraphics[clip=true,width=7cm]{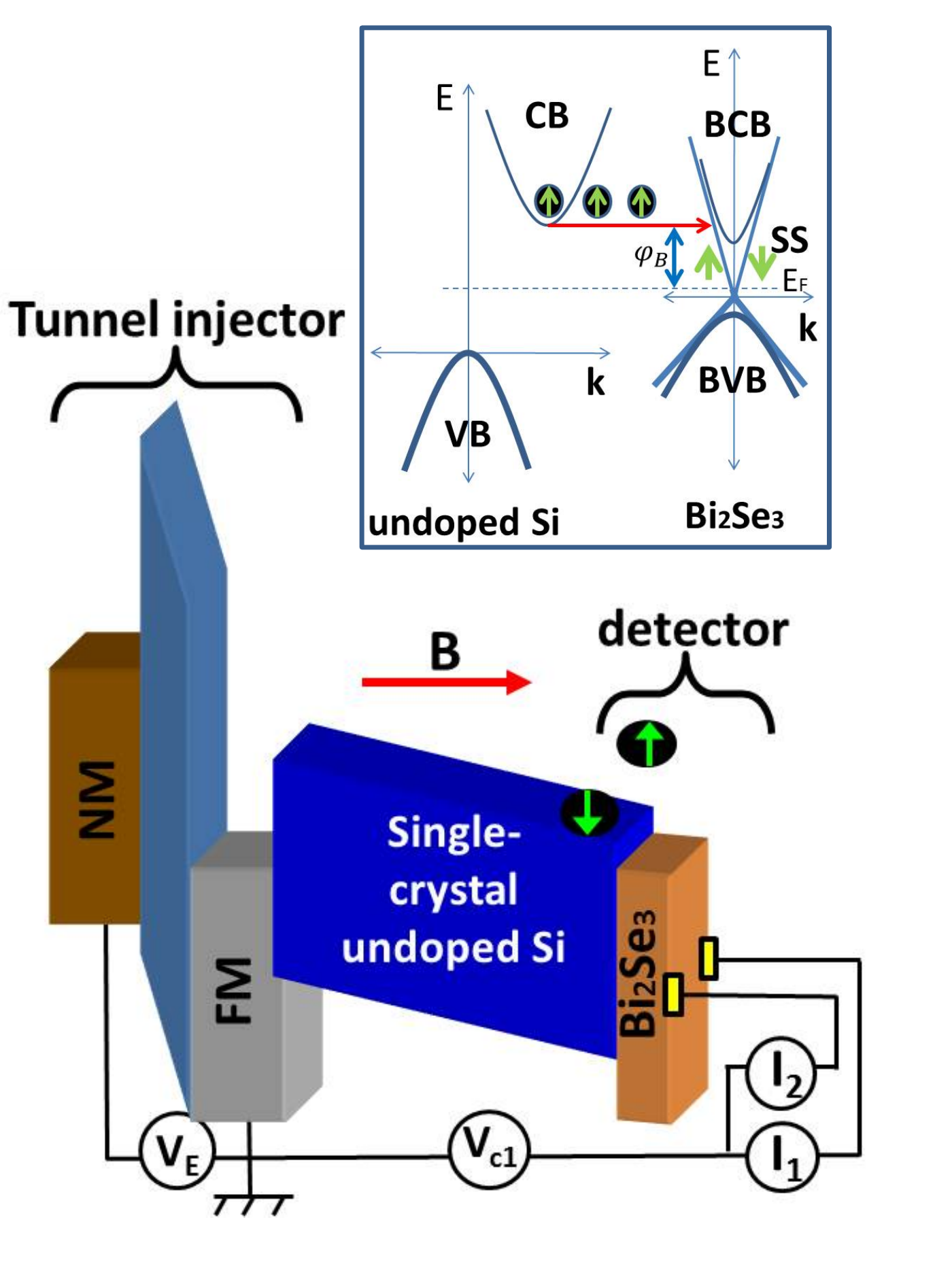}
    \caption{Diagram of the proposed scheme to inject a spin polarized current into a topological insulator. A tunnel junction injects unpolarized hot electrons from a nonmagnetic metal (NM) cathode into a ferromagnetic anode (FM). After attenuation of minority spin electrons, electrons enter the Si conduction band accelerated by a voltage V$_{c1}$, and enter the Bi$_2$Se$_3$ with excess energy given by the Bi$_2$Se$_3$-Si Schottky barrier height. Spin-momentum conversion in Bi$_2$Se$_3$ leads to a spin dependent difference in currents I$_1$ and I$_2$. The direction of injected spin is controlled by precession in the applied magnetic field during the Si transport time $\bar{t}$ (see text). Inset: schematic of band structure of Si and Bi$_2$Se$_3$ showing spin polarized surface states (SS) and unpolarized bulk conduction (BCB) and valence (BVB) bands of Bi$_2$Se$_3$.}
    \label{Scheme}
\end{figure}

As first steps towards this goal, we have demonstrated the feasibility of fabricating mesoscopic Bi$_2$Se$_3$ devices on bare silicon, measured charge injection across the silicon-Bi$_2$Se$_3$ interface, and determined the Schottky barrier height of 0.34 eV. This discontinuity between Si conduction band and TI Fermi level determines the initial electron state within the TI bandstructure upon injection across the interface from the Si channel.

We first describe the proposed experiment to use spin injection from silicon to measure the spin Hall effect in Bi$_2$Se$_3$. As shown in the scheme in Figure \ref{Scheme}, unpolarized hot electrons produced by a tunnel junction are injected from a nonmagnetic cathode (NM) into a ferromagnetic anode (FM) where spin-dependent electron scattering attenuates minority spin electrons. This mechanism has been shown to produce spin polarization over 90\% in the ballistic component of this current crossing the Schottky barrier with Si on the other side \cite{Monsma_PRL1995, vanDijken_APL2003}. After relaxation to the Si conduction band minimum, these spin polarized hot electrons can cross the full thickness of a Si wafer (several hundred microns) under influence of an applied electric field. In Ref. \onlinecite{Huang_PRL2007}, a subsequent FM spin filter was used to make a spin-dependent measurement of final spin angle and polarization demonstrating coherent spin precession at frequency $\omega=g\mu_BB/\hbar$ during transport through 350 micron-thick Si at temperatures up to 150K. The proposed experiment here will detect the spin through conversion to a charge current through the spin Hall effect due to spin-momentum locking in a TI as follows.

After transport through the Si, spin polarized current is injected into the TI which is electrically contacted at two opposite sides. The injected current in the proposed scheme (where the ferromagnet is placed in the anode of the tunnel junction) can reach 100nA \cite{Appelbaum_Nature2007}. If spin is locked to momentum, spin ``up'' will preferentially flow toward one contact, whereas spin ``down'' will flow to the other, constituting an intrinsic spin-Hall effect. A small perpendicular magnetic field will cause the spins to precess while transiting the silicon, controlling the final spin orientation at the Si-TI interface. The spin angle is $g \mu_B B \bar{t} / \hbar$ where $\hbar$ is the reduced Planck constant, $\mu_B$ is the Bohr magneton, $B$ the magnetic field and $g$ the gyromagnetic ratio. For mean transit time $\bar{t}\approx$10~ns, precession through a full period $2\pi$ requires application of $\approx$36 gauss, not enough to significantly impact the gapless surface states.

Due to spin-momentum locking, the effect is ballistic in nature. The spin polarization of the arriving electrons is detectable only in a region within one mean free path around each electrode, since the mean free path is identical to the spin relaxation length. Hence the signal is expected to scale as $l/L$ where $l$ is the electron mean free path and $L$ is the sample dimension. To maximize the signal the electrodes contacting the TI crystal should be as closely spaced as possible.

\begin{figure}
    \includegraphics[clip=true,width=9cm]{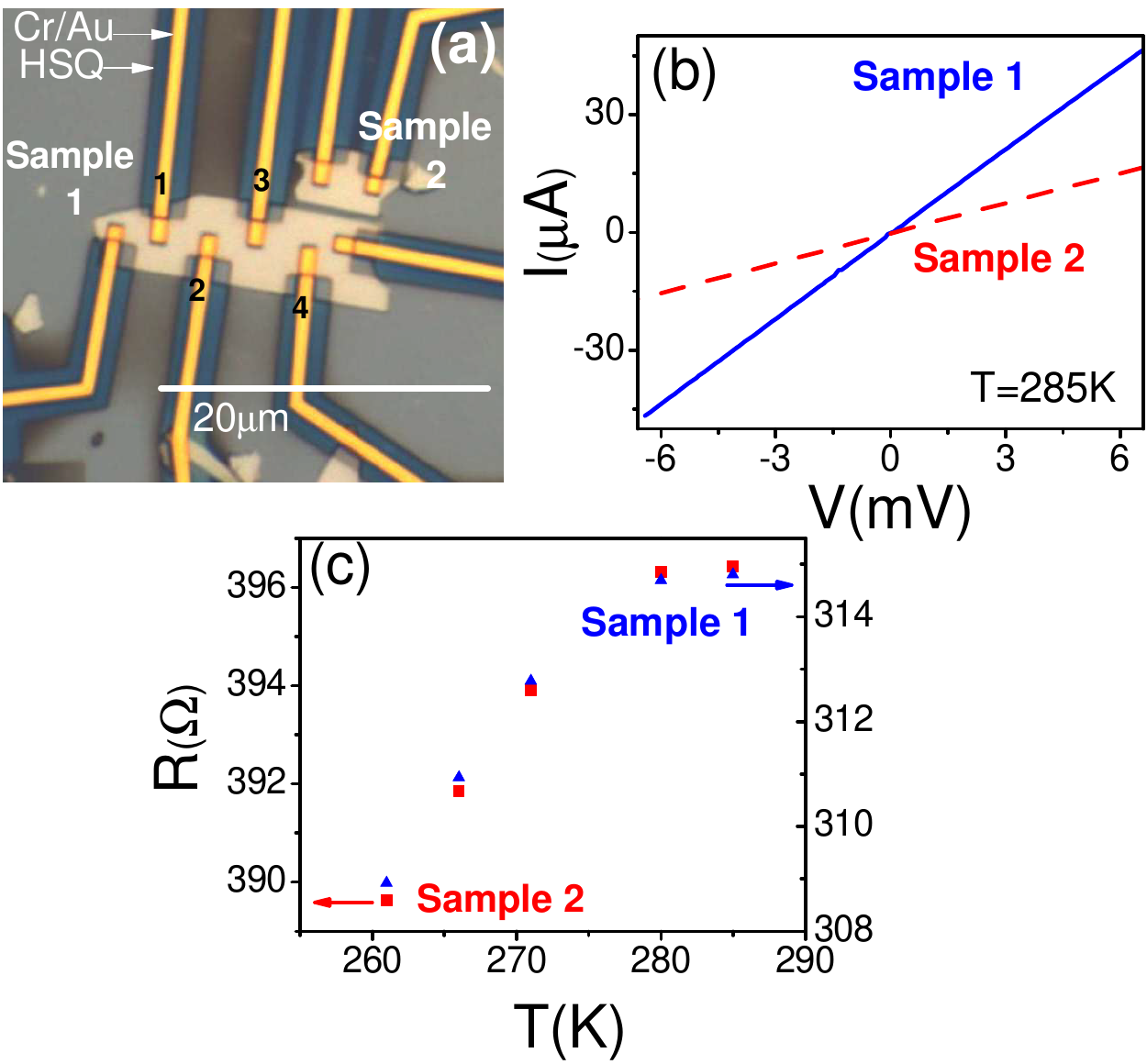}
    \caption{Transport measurements of two of the Bi$_2$Se$_3$ crystals studied. (a) Optical image of Bi$_2$Se$_3$ samples showing Cr/Au electrodes and HSQ isolation layer to prevent direct contact between Cr/Au electrodes and Si. (b) Linear four probe and two probe I-V characteristics of samples 1 and 2 respectively. (In sample 1 current was driven through electrodes 1 and 4 and voltage was measured in electrodes 2 and 3). (c) Two probe resistance of samples 1 and 2 as a function of temperature.}
    \label{Fig1}
\end{figure}

For successful injection of spin polarized electrons into the surface states within the bandgap of the TI, the Si conduction band-TI Dirac point discontinuity should be minimized. ARPES experiments indicate that the TI surface state persists as a resonance to energies above the conduction band edge, and retains a degree of spin polarization \cite{Pan_PRL2011,Gedik2011}. However, injection at energies higher than the bulk conduction band edge will populate the spin-textured surface state resonance as well as the spin degenerate bulk states, reducing the signal. Inelastic processes which relax the hot carriers may also relax the injected spin and further reduce the signal. It is therefore of great importance to know the Si conduction band-TI Fermi level discontinuity (Schottky barrier). In the following we describe the procedure we have followed to measure experimentally the Si-Bi$_2$Se$_3$ Schottky barrier.

Exfoliated crystals of Bi$_2$Se$_3$ were deposited on n-type low doped silicon (1-10$\Omega$ cm) and a technique was developed as follows to fabricate electrodes on the crystals without having an electrical contact to the silicon. Crystals were grown as indicated in Ref. \onlinecite{Butch_PRB2010} with a bulk charge density of $\approx 10^{17}$ cm$^{-3}$. Prior to exfoliation, the silicon wafer was immersed in $2\%$ buffered hydrofluoric acid (BHF) to remove any oxide layer formed and obtain a good Schottky contact. After exfoliation, thin crystals of about $70$~nm thick were identified optically. A negative resist (hydrogen silsesquioxane, HSQ) was deposited and patterned using electron-beam lithography as an insulating layer between the electrodes and silicon. Electrodes were patterned with a positive resist using electron-beam lithography. A good electrical contact to the TI is achieved using a N$_2$ plasma for several seconds in the contact area \cite{Dohun_NatPhys2012} before depositing the Cr/Au (10nm/100nm) electrodes.

Figure \ref{Fig1}(a) shows an optical image of the sample consisting of two independent Bi$_2$Se$_3$ crystals, sample 1 and sample 2. Four probe resistance measurements on sample 1 and two probe measurements on sample 2 (Fig. \ref{Fig1}(b)) show linear Ohmic behavior, indicating a good electrical contact to the Bi$_2$Se$_3$. The sheet resistance of the Bi$_2$Se$_3$ was found to be 140$\Omega$ and contact resistance $176\Omega$. The increasing resistance of samples 1 and 2 with increasing temperature shown in Fig. \ref{Fig1}(c) indicates metallic behavior.

\begin{figure}
    \includegraphics[clip=true,width=9.1cm]{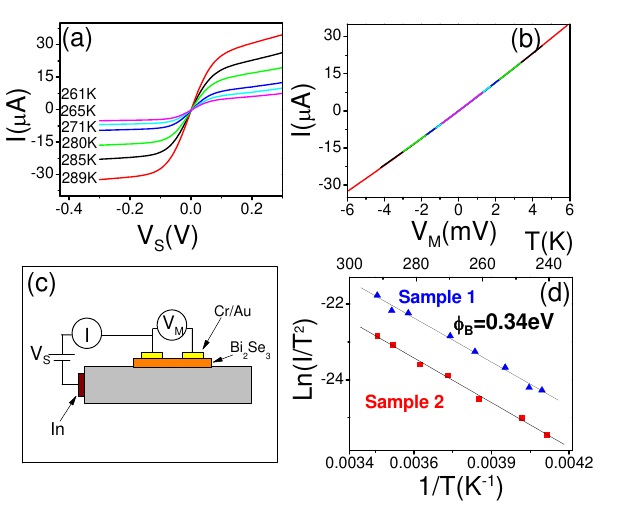}
    \caption{(a-b) Current I through the In-silicon-Bi$_2$Se$_3$-Cr/Au structure as a function of $V_S$ measured between In and Cr/Au and a second contact on the Bi$_2$Se$_3$ shown in (b). (c) shows a schematic of the measurement setup. Measurements at different temperatures T are shown as indicated by legend in (a). (c) Semi-log plot of $I/T^2$ as a function of $1/T$ for Si-Bi$_2$Se$_3$ junctions sample 1 and sample 2. Straight lines are fits to the Richardson-Dushman thermionic emission theory (Equation 1) with a Schottky barrier height of 0.34eV.}
    \label{Fig2}
\end{figure}

Figure \ref{Fig2}(a) shows the I-V characteristic measured between an indium contact to Si and electrode 3 on Bi$_2$Se$_3$ sample 1 at
different temperatures. We interpret the sublinear I-V characteristic as that of two back-to-back Schottky diodes: one at the Bi$_2$Se$_3$-Si interface and one at the indium-Si interface, such that one of the two is in reverse bias regardless of voltage polarity. Figure \ref{Fig2}(b) shows the same current vs $V_M$, the voltage drop measured across the Bi$_2$Se$_3$-Cr/Au interface, as it is represented in figure \ref{Fig2}(c). Here Ohmic behavior indicates that the sublinear characteristic in figure \ref{Fig2}(a) does not come from the Bi$_2$Se$_3$-Cr/Au contact.
The Si-Bi$_2$Se$_3$ Schottky barrier height was deduced from the reverse bias saturation current. The negative side of the I-V curve reflects the saturation of the Bi$_2$Se$_3$-Si diode and the positive side the saturation of the indium-Si diode. Figure \ref{Fig2}(d) shows the fit of the saturation current of the Bi$_2$Se$_3$-Si diode with the Richardson-Dushman thermionic emission theory,

\be \label{Emission}
J=-A^*T^2\exp\Big(-\frac{q\phi_{B}}{k_BT}\Big),
\ee

\noindent where $\phi_B$ is the Schottky barrier height and $A^*$ the Richardson constant, 4$\pi$ qm$^*$k$_B^2$/h$^3$. The deduced values were $\phi_B=0.34\pm0.01$ eV for sample 1 and $\phi_B=0.34\pm0.009$ eV for sample 2. The same value has been found on a third Bi$_2$Se$_3$ crystal 5 times smaller than sample 2.

The Richardson constant in samples 1 and 2 is approximately 300A/cm$^2$/K$^2$, three times larger than what is expected theoretically for Si \cite{Sze_2007}. Such large values for the Richardson constant have previously been observed in Si diodes with different metals, due to a strong sensitivity to interfacial states \cite{Toyama_APL1985, Toyama_JAP1988}.

The measured small Si-Bi$_2$Se$_3$ Schottky barrier height of 0.34eV is reasonable given that the ionization potential of n-type Bi$_2$Se$_3$ is $\approx$4.45 eV \cite{PersonalHsiehGedik} and electron affinity of Si is 4.05 eV. (Our Bi$_2$Se$_3$ samples after exfoliation have a Fermi energy placed near the bottom of the bulk conduction band \cite{Sungjae_Nano2011,Dohun_NatPhys2012}). Because the bandgap of Bi$_2$Se$_3$ is only 0.3 eV $<$ $\phi_B$ \cite{Xia_NatPhys2009}, electrons from the Si conduction band are resonant with bulk conduction band states in the Bi$_2$Se$_3$ even if the Fermi level lies in the gap, hence injection will occur into the spin-textured surface resonance as well as the spin degenerate bulk states. This difficulty can be potentially addressed by effectively eliminating the Schottky barrier by degenerately doping the silicon surface such that the depletion region is transparent to tunneling.

Alternatively, the existence of a robust Schottky barrier at the TI-Si interface can be exploited in other novel device designs intended for study of the spin-polarized properties of the surface states. For example, oblique illumination of the TI-Si interface with circularly-polarized infrared photons of energy $h\nu>$0.34eV will result in spin-selective surface-state electron excitation into the Si conduction band, in an \emph{internal} process \cite{SVPD_Appelbaum2003} analogous to recent (vacuum) photoemission experiments \cite{Wang_PRL2011}, without the need for high-energy ultraviolet photons. Although energy and momentum dispersion information is lost by scattering during transport through the Si, subsequent spin detection of this photocurrent with semiconductor-FM metal-semiconductor structures as in conventional spintronic devices \cite{Appelbaum_Nature2007} can then be used to study the spin properties of the surface state assuming $h\nu<\approx 0.7$eV (set by the Schottky barrier height of the FM/Si detector interface). This scheme has the added benefit that external probes such as mechanical strain (precluded by the requirements of surface science methods like ARPES) can be readily applied in the ex-situ measurement environment.

This work has been supported by the University of Maryland NSF-MRSEC and MRSEC shared facilities under Grant No. DMR 05-20741, NSF grant DMR 11-05224 and the Office of Naval Research. We acknowledge the support of the Maryland NanoCenter and its FabLab in particular Tom Loughran, and helpful discussions with Dohun Kim.

\bibliography{SchottkyTI-V9-Arxiv}

\begin{thebibliography}{27}%
\makeatletter
\providecommand \@ifxundefined [1]{%
 \@ifx{#1\undefined}
}%
\providecommand \@ifnum [1]{%
 \ifnum #1\expandafter \@firstoftwo
 \else \expandafter \@secondoftwo
 \fi
}%
\providecommand \@ifx [1]{%
 \ifx #1\expandafter \@firstoftwo
 \else \expandafter \@secondoftwo
 \fi
}%
\providecommand \natexlab [1]{#1}%
\providecommand \enquote  [1]{``#1''}%
\providecommand \bibnamefont  [1]{#1}%
\providecommand \bibfnamefont [1]{#1}%
\providecommand \citenamefont [1]{#1}%
\providecommand \href@noop [0]{\@secondoftwo}%
\providecommand \href [0]{\begingroup \@sanitize@url \@href}%
\providecommand \@href[1]{\@@startlink{#1}\@@href}%
\providecommand \@@href[1]{\endgroup#1\@@endlink}%
\providecommand \@sanitize@url [0]{\catcode `\\12\catcode `\$12\catcode
  `\&12\catcode `\#12\catcode `\^12\catcode `\_12\catcode `\%12\relax}%
\providecommand \@@startlink[1]{}%
\providecommand \@@endlink[0]{}%
\providecommand \url  [0]{\begingroup\@sanitize@url \@url }%
\providecommand \@url [1]{\endgroup\@href {#1}{\urlprefix }}%
\providecommand \urlprefix  [0]{URL }%
\providecommand \Eprint [0]{\href }%
\providecommand \doibase [0]{http://dx.doi.org/}%
\providecommand \selectlanguage [0]{\@gobble}%
\providecommand \bibinfo  [0]{\@secondoftwo}%
\providecommand \bibfield  [0]{\@secondoftwo}%
\providecommand \translation [1]{[#1]}%
\providecommand \BibitemOpen [0]{}%
\providecommand \bibitemStop [0]{}%
\providecommand \bibitemNoStop [0]{.\EOS\space}%
\providecommand \EOS [0]{\spacefactor3000\relax}%
\providecommand \BibitemShut  [1]{\csname bibitem#1\endcsname}%
\let\auto@bib@innerbib\@empty
%</preamble>
\bibitem [{\citenamefont {Xia}\ \emph {et~al.}(2009)\citenamefont {Xia},
  \citenamefont {Qian}, \citenamefont {Hsieh}, \citenamefont {Wray},
  \citenamefont {Pal}, \citenamefont {Lin}, \citenamefont {Bansil},
  \citenamefont {Grauer}, \citenamefont {Cava},\ and\ \citenamefont
  {Hasan}}]{Xia_NatPhys2009}%
  \BibitemOpen
  \bibfield  {author} {\bibinfo {author} {\bibfnamefont {Y.}~\bibnamefont
  {Xia}}, \bibinfo {author} {\bibfnamefont {D.}~\bibnamefont {Qian}}, \bibinfo
  {author} {\bibfnamefont {D.}~\bibnamefont {Hsieh}}, \bibinfo {author}
  {\bibfnamefont {L.}~\bibnamefont {Wray}}, \bibinfo {author} {\bibfnamefont
  {A.}~\bibnamefont {Pal}}, \bibinfo {author} {\bibfnamefont {H.}~\bibnamefont
  {Lin}}, \bibinfo {author} {\bibfnamefont {A.}~\bibnamefont {Bansil}},
  \bibinfo {author} {\bibfnamefont {D.}~\bibnamefont {Grauer}}, \bibinfo
  {author} {\bibfnamefont {R.~J.}\ \bibnamefont {Cava}}, \ and\ \bibinfo
  {author} {\bibfnamefont {M.~Z.}\ \bibnamefont {Hasan}},\ }\href@noop {}
  {\bibfield  {journal} {\bibinfo  {journal} {Nat. Phys.}\ }\textbf {\bibinfo
  {volume} {5}},\ \bibinfo {pages} {398} (\bibinfo {year} {2009})}\BibitemShut
  {NoStop}%
\bibitem [{\citenamefont {Hasan}\ \emph {et~al.}(2011)\citenamefont {Hasan},
  \citenamefont {Hsieh}, \citenamefont {Xia}, \citenamefont {Wray},
  \citenamefont {Xu},\ and\ \citenamefont {Kane}}]{Hasan_Rev2012}%
  \BibitemOpen
  \bibfield  {author} {\bibinfo {author} {\bibfnamefont {M.~Z.}\ \bibnamefont
  {Hasan}}, \bibinfo {author} {\bibfnamefont {D.}~\bibnamefont {Hsieh}},
  \bibinfo {author} {\bibfnamefont {Y.}~\bibnamefont {Xia}}, \bibinfo {author}
  {\bibfnamefont {A.~L.}\ \bibnamefont {Wray}}, \bibinfo {author}
  {\bibfnamefont {S.~Y.}\ \bibnamefont {Xu}}, \ and\ \bibinfo {author}
  {\bibfnamefont {C.~L.}\ \bibnamefont {Kane}},\ }\href@noop {} {\bibfield
  {journal} {\bibinfo  {journal} {arXiv: 1105.0396}\ } (\bibinfo {year}
  {2011})}\BibitemShut {NoStop}%
\bibitem [{\citenamefont {Pan}\ \emph {et~al.}(2011)\citenamefont {Pan},
  \citenamefont {Vescovo}, \citenamefont {Fedorov}, \citenamefont {Gardner},
  \citenamefont {Lee}, \citenamefont {Chu}, \citenamefont {Gu},\ and\
  \citenamefont {Valla}}]{Pan_PRL2011}%
  \BibitemOpen
  \bibfield  {author} {\bibinfo {author} {\bibfnamefont {Z.-H.}\ \bibnamefont
  {Pan}}, \bibinfo {author} {\bibfnamefont {E.}~\bibnamefont {Vescovo}},
  \bibinfo {author} {\bibfnamefont {A.~V.}\ \bibnamefont {Fedorov}}, \bibinfo
  {author} {\bibfnamefont {D.}~\bibnamefont {Gardner}}, \bibinfo {author}
  {\bibfnamefont {Y.~S.}\ \bibnamefont {Lee}}, \bibinfo {author} {\bibfnamefont
  {S.}~\bibnamefont {Chu}}, \bibinfo {author} {\bibfnamefont {G.~D.}\
  \bibnamefont {Gu}}, \ and\ \bibinfo {author} {\bibfnamefont {T.}~\bibnamefont
  {Valla}},\ }\href@noop {} {\bibfield  {journal} {\bibinfo  {journal} {Phys.
  Rev. Lett.}\ }\textbf {\bibinfo {volume} {106}},\ \bibinfo {pages} {257004}
  (\bibinfo {year} {2011})}\BibitemShut {NoStop}%
\bibitem [{\citenamefont {Hsieh}\ \emph {et~al.}(2009)\citenamefont {Hsieh},
  \citenamefont {Xia}, \citenamefont {Qian}, \citenamefont {Wray},
  \citenamefont {Dil}, \citenamefont {Meier}, \citenamefont {Osterwalder},
  \citenamefont {Patthey}, \citenamefont {Checkelsky}, \citenamefont {Ong},
  \citenamefont {Fedorov}, \citenamefont {Lin}, \citenamefont {Bansil},
  \citenamefont {Grauer}, \citenamefont {Hor}, \citenamefont {Cava},\ and\
  \citenamefont {Hasan}}]{Hsieh_Nature2009}%
  \BibitemOpen
  \bibfield  {author} {\bibinfo {author} {\bibfnamefont {D.}~\bibnamefont
  {Hsieh}}, \bibinfo {author} {\bibfnamefont {Y.}~\bibnamefont {Xia}}, \bibinfo
  {author} {\bibfnamefont {D.}~\bibnamefont {Qian}}, \bibinfo {author}
  {\bibfnamefont {L.}~\bibnamefont {Wray}}, \bibinfo {author} {\bibfnamefont
  {J.~H.}\ \bibnamefont {Dil}}, \bibinfo {author} {\bibfnamefont
  {F.}~\bibnamefont {Meier}}, \bibinfo {author} {\bibfnamefont
  {J.}~\bibnamefont {Osterwalder}}, \bibinfo {author} {\bibfnamefont
  {L.}~\bibnamefont {Patthey}}, \bibinfo {author} {\bibfnamefont {J.~G.}\
  \bibnamefont {Checkelsky}}, \bibinfo {author} {\bibfnamefont {N.~P.}\
  \bibnamefont {Ong}}, \bibinfo {author} {\bibfnamefont {A.~V.}\ \bibnamefont
  {Fedorov}}, \bibinfo {author} {\bibfnamefont {H.}~\bibnamefont {Lin}},
  \bibinfo {author} {\bibfnamefont {A.}~\bibnamefont {Bansil}}, \bibinfo
  {author} {\bibfnamefont {D.}~\bibnamefont {Grauer}}, \bibinfo {author}
  {\bibfnamefont {Y.~S.}\ \bibnamefont {Hor}}, \bibinfo {author} {\bibfnamefont
  {R.~J.}\ \bibnamefont {Cava}}, \ and\ \bibinfo {author} {\bibfnamefont
  {M.~Z.}\ \bibnamefont {Hasan}},\ }\href@noop {} {\bibfield  {journal}
  {\bibinfo  {journal} {Nature}\ }\textbf {\bibinfo {volume} {460}},\ \bibinfo
  {pages} {1101} (\bibinfo {year} {2009})}\BibitemShut {NoStop}%
\bibitem [{\citenamefont {Wang}\ \emph
  {et~al.}(2011{\natexlab{a}})\citenamefont {Wang}, \citenamefont {Hsieh},
  \citenamefont {Pilon}, \citenamefont {Fu}, \citenamefont {Gardner},
  \citenamefont {Lee},\ and\ \citenamefont {Gedik}}]{Wang_PRL2011}%
  \BibitemOpen
  \bibfield  {author} {\bibinfo {author} {\bibfnamefont {Y.~H.}\ \bibnamefont
  {Wang}}, \bibinfo {author} {\bibfnamefont {D.}~\bibnamefont {Hsieh}},
  \bibinfo {author} {\bibfnamefont {D.}~\bibnamefont {Pilon}}, \bibinfo
  {author} {\bibfnamefont {L.}~\bibnamefont {Fu}}, \bibinfo {author}
  {\bibfnamefont {D.~R.}\ \bibnamefont {Gardner}}, \bibinfo {author}
  {\bibfnamefont {Y.~S.}\ \bibnamefont {Lee}}, \ and\ \bibinfo {author}
  {\bibfnamefont {N.}~\bibnamefont {Gedik}},\ }\href@noop {} {\bibfield
  {journal} {\bibinfo  {journal} {Phys. Rev. Lett.}\ }\textbf {\bibinfo
  {volume} {107}},\ \bibinfo {pages} {207602} (\bibinfo {year}
  {2011}{\natexlab{a}})}\BibitemShut {NoStop}%
\bibitem [{\citenamefont {Lou}\ \emph {et~al.}(2007)\citenamefont {Lou},
  \citenamefont {Adelmann}, \citenamefont {Crooker}, \citenamefont {Garlid},
  \citenamefont {Zhang}, \citenamefont {Reddy}, \citenamefont {Flexner},
  \citenamefont {Palmstr{\o}m},\ and\ \citenamefont
  {Crowell}}]{Lou_NatPhys2007}%
  \BibitemOpen
  \bibfield  {author} {\bibinfo {author} {\bibfnamefont {X.}~\bibnamefont
  {Lou}}, \bibinfo {author} {\bibfnamefont {C.}~\bibnamefont {Adelmann}},
  \bibinfo {author} {\bibfnamefont {S.}~\bibnamefont {Crooker}}, \bibinfo
  {author} {\bibfnamefont {E.}~\bibnamefont {Garlid}}, \bibinfo {author}
  {\bibfnamefont {J.}~\bibnamefont {Zhang}}, \bibinfo {author} {\bibfnamefont
  {S.}~\bibnamefont {Reddy}}, \bibinfo {author} {\bibfnamefont
  {S.}~\bibnamefont {Flexner}}, \bibinfo {author} {\bibfnamefont
  {C.}~\bibnamefont {Palmstr{\o}m}}, \ and\ \bibinfo {author} {\bibfnamefont
  {P.~A.}\ \bibnamefont {Crowell}},\ }\href@noop {} {\bibfield  {journal}
  {\bibinfo  {journal} {Nature Phys.}\ }\textbf {\bibinfo {volume} {3}},\
  \bibinfo {pages} {197} (\bibinfo {year} {2007})}\BibitemShut {NoStop}%
\bibitem [{\citenamefont {Sasaki}\ \emph {et~al.}(2010)\citenamefont {Sasaki},
  \citenamefont {Oikawa}, \citenamefont {Suzuki}, \citenamefont {Shiraishi},
  \citenamefont {Suzuki},\ and\ \citenamefont {Noguchi}}]{Sasaki_APL2010}%
  \BibitemOpen
  \bibfield  {author} {\bibinfo {author} {\bibfnamefont {T.}~\bibnamefont
  {Sasaki}}, \bibinfo {author} {\bibfnamefont {T.}~\bibnamefont {Oikawa}},
  \bibinfo {author} {\bibfnamefont {T.}~\bibnamefont {Suzuki}}, \bibinfo
  {author} {\bibfnamefont {M.}~\bibnamefont {Shiraishi}}, \bibinfo {author}
  {\bibfnamefont {Y.}~\bibnamefont {Suzuki}}, \ and\ \bibinfo {author}
  {\bibfnamefont {K.}~\bibnamefont {Noguchi}},\ }\href@noop {} {\bibfield
  {journal} {\bibinfo  {journal} {Appl. Phys. Lett.}\ }\textbf {\bibinfo
  {volume} {96}},\ \bibinfo {pages} {122101} (\bibinfo {year}
  {2010})}\BibitemShut {NoStop}%
\bibitem [{\citenamefont {Appelbaum}, \citenamefont {Huang},\ and\
  \citenamefont {Monsma}(2007)}]{Appelbaum_Nature2007}%
  \BibitemOpen
  \bibfield  {author} {\bibinfo {author} {\bibfnamefont {I.}~\bibnamefont
  {Appelbaum}}, \bibinfo {author} {\bibfnamefont {B.}~\bibnamefont {Huang}}, \
  and\ \bibinfo {author} {\bibfnamefont {D.~J.}\ \bibnamefont {Monsma}},\
  }\href@noop {} {\bibfield  {journal} {\bibinfo  {journal} {Nature}\ }\textbf
  {\bibinfo {volume} {447}},\ \bibinfo {pages} {295} (\bibinfo {year}
  {2007})}\BibitemShut {NoStop}%
\bibitem [{\citenamefont {Johnson}\ and\ \citenamefont
  {Silsbee}(1985)}]{Johnson_PRL1985}%
  \BibitemOpen
  \bibfield  {author} {\bibinfo {author} {\bibfnamefont {M.}~\bibnamefont
  {Johnson}}\ and\ \bibinfo {author} {\bibfnamefont {R.~H.}\ \bibnamefont
  {Silsbee}},\ }\href@noop {} {\bibfield  {journal} {\bibinfo  {journal} {Phys.
  Rev. Lett.}\ }\textbf {\bibinfo {volume} {55}},\ \bibinfo {pages} {1790}
  (\bibinfo {year} {1985})}\BibitemShut {NoStop}%
\bibitem [{\citenamefont {Goldberg}\ and\ \citenamefont
  {Davis}(1954)}]{Goldberg_PR1954}%
  \BibitemOpen
  \bibfield  {author} {\bibinfo {author} {\bibfnamefont {C.}~\bibnamefont
  {Goldberg}}\ and\ \bibinfo {author} {\bibfnamefont {R.~E.}\ \bibnamefont
  {Davis}},\ }\href@noop {} {\bibfield  {journal} {\bibinfo  {journal} {Phys.
  Rev.}\ }\textbf {\bibinfo {volume} {94}},\ \bibinfo {pages} {1121} (\bibinfo
  {year} {1954})}\BibitemShut {NoStop}%
\bibitem [{\citenamefont {Bullis}\ and\ \citenamefont
  {Krag}(1956)}]{Bullis_PR1956}%
  \BibitemOpen
  \bibfield  {author} {\bibinfo {author} {\bibfnamefont {W.~M.}\ \bibnamefont
  {Bullis}}\ and\ \bibinfo {author} {\bibfnamefont {W.~E.}\ \bibnamefont
  {Krag}},\ }\href@noop {} {\bibfield  {journal} {\bibinfo  {journal} {Phys.
  Rev.}\ }\textbf {\bibinfo {volume} {101}},\ \bibinfo {pages} {580} (\bibinfo
  {year} {1956})}\BibitemShut {NoStop}%
\bibitem [{\citenamefont {Campbell}\ and\ \citenamefont
  {Fert}(1982)}]{CampbellFert_1982}%
  \BibitemOpen
  \bibfield  {author} {\bibinfo {author} {\bibfnamefont {I.~A.}\ \bibnamefont
  {Campbell}}\ and\ \bibinfo {author} {\bibfnamefont {A.}~\bibnamefont
  {Fert}},\ }\href@noop {} {\emph {\bibinfo {title} {Ferromagnetic
  materials}}}\ (\bibinfo  {publisher} {North-Holland, Amsterdam},\ \bibinfo
  {year} {1982})\BibitemShut {NoStop}%
\bibitem [{\citenamefont {Xia}\ \emph {et~al.}(2011)\citenamefont {Xia},
  \citenamefont {Li}, \citenamefont {Ke}, \citenamefont {Liu}, \citenamefont
  {Ren}, \citenamefont {Su}, \citenamefont {Huan}, \citenamefont {Dong},\ and\
  \citenamefont {Wang}}]{Xia_Arxiv2012}%
  \BibitemOpen
  \bibfield  {author} {\bibinfo {author} {\bibfnamefont {B.}~\bibnamefont
  {Xia}}, \bibinfo {author} {\bibfnamefont {Z.~P.}\ \bibnamefont {Li}},
  \bibinfo {author} {\bibfnamefont {C.}~\bibnamefont {Ke}}, \bibinfo {author}
  {\bibfnamefont {P.}~\bibnamefont {Liu}}, \bibinfo {author} {\bibfnamefont
  {P.}~\bibnamefont {Ren}}, \bibinfo {author} {\bibfnamefont {H.~B.}\
  \bibnamefont {Su}}, \bibinfo {author} {\bibfnamefont {A.~C.~H.}\ \bibnamefont
  {Huan}}, \bibinfo {author} {\bibfnamefont {Z.~L.}\ \bibnamefont {Dong}}, \
  and\ \bibinfo {author} {\bibfnamefont {L.}~\bibnamefont {Wang}},\ }\href@noop
  {} {\bibfield  {journal} {\bibinfo  {journal} {arXiv:1109.1379}\ } (\bibinfo
  {year} {2011})}\BibitemShut {NoStop}%
\bibitem [{\citenamefont {Moser}\ \emph {et~al.}(2007)\citenamefont {Moser},
  \citenamefont {Matos-Abiague}, \citenamefont {Schuh}, \citenamefont
  {Wegscheider}, \citenamefont {Fabian},\ and\ \citenamefont
  {Weiss}}]{Moser_PRL2007}%
  \BibitemOpen
  \bibfield  {author} {\bibinfo {author} {\bibfnamefont {J.}~\bibnamefont
  {Moser}}, \bibinfo {author} {\bibfnamefont {A.}~\bibnamefont
  {Matos-Abiague}}, \bibinfo {author} {\bibfnamefont {D.}~\bibnamefont
  {Schuh}}, \bibinfo {author} {\bibfnamefont {W.}~\bibnamefont {Wegscheider}},
  \bibinfo {author} {\bibfnamefont {J.}~\bibnamefont {Fabian}}, \ and\ \bibinfo
  {author} {\bibfnamefont {D.}~\bibnamefont {Weiss}},\ }\href@noop {}
  {\bibfield  {journal} {\bibinfo  {journal} {Phys. Rev. Lett.}\ }\textbf
  {\bibinfo {volume} {99}},\ \bibinfo {pages} {056601} (\bibinfo {year}
  {2007})}\BibitemShut {NoStop}%
\bibitem [{\citenamefont {Uemura}\ \emph {et~al.}(2011)\citenamefont {Uemura},
  \citenamefont {Harada}, \citenamefont {Akiho}, \citenamefont {ichi Matsuda},\
  and\ \citenamefont {Yamamoto}}]{Uemura_APL2011}%
  \BibitemOpen
  \bibfield  {author} {\bibinfo {author} {\bibfnamefont {T.}~\bibnamefont
  {Uemura}}, \bibinfo {author} {\bibfnamefont {M.}~\bibnamefont {Harada}},
  \bibinfo {author} {\bibfnamefont {T.}~\bibnamefont {Akiho}}, \bibinfo
  {author} {\bibfnamefont {K.}~\bibnamefont {ichi Matsuda}}, \ and\ \bibinfo
  {author} {\bibfnamefont {M.}~\bibnamefont {Yamamoto}},\ }\href@noop {}
  {\bibfield  {journal} {\bibinfo  {journal} {Applied Physics Letters}\
  }\textbf {\bibinfo {volume} {98}},\ \bibinfo {pages} {102503} (\bibinfo
  {year} {2011})}\BibitemShut {NoStop}%
\bibitem [{\citenamefont {Monsma}\ \emph {et~al.}(1995)\citenamefont {Monsma},
  \citenamefont {Lodder}, \citenamefont {Popma},\ and\ \citenamefont
  {Dieny}}]{Monsma_PRL1995}%
  \BibitemOpen
  \bibfield  {author} {\bibinfo {author} {\bibfnamefont {D.~J.}\ \bibnamefont
  {Monsma}}, \bibinfo {author} {\bibfnamefont {J.~C.}\ \bibnamefont {Lodder}},
  \bibinfo {author} {\bibfnamefont {T.~J.~A.}\ \bibnamefont {Popma}}, \ and\
  \bibinfo {author} {\bibfnamefont {B.}~\bibnamefont {Dieny}},\ }\href@noop {}
  {\bibfield  {journal} {\bibinfo  {journal} {Phys. Rev. Lett.}\ }\textbf
  {\bibinfo {volume} {74}},\ \bibinfo {pages} {5260} (\bibinfo {year}
  {1995})}\BibitemShut {NoStop}%
\bibitem [{\citenamefont {van Dijken}, \citenamefont {Jiang},\ and\
  \citenamefont {Parkin}(2003)}]{vanDijken_APL2003}%
  \BibitemOpen
  \bibfield  {author} {\bibinfo {author} {\bibfnamefont {S.}~\bibnamefont {van
  Dijken}}, \bibinfo {author} {\bibfnamefont {X.}~\bibnamefont {Jiang}}, \ and\
  \bibinfo {author} {\bibfnamefont {S.~S.~P.}\ \bibnamefont {Parkin}},\
  }\href@noop {} {\bibfield  {journal} {\bibinfo  {journal} {Applied Physics
  Letters}\ }\textbf {\bibinfo {volume} {83}},\ \bibinfo {pages} {951}
  (\bibinfo {year} {2003})}\BibitemShut {NoStop}%
\bibitem [{\citenamefont {Huang}, \citenamefont {Monsma},\ and\ \citenamefont
  {Appelbaum}(2007)}]{Huang_PRL2007}%
  \BibitemOpen
  \bibfield  {author} {\bibinfo {author} {\bibfnamefont {B.}~\bibnamefont
  {Huang}}, \bibinfo {author} {\bibfnamefont {D.~J.}\ \bibnamefont {Monsma}}, \
  and\ \bibinfo {author} {\bibfnamefont {I.}~\bibnamefont {Appelbaum}},\
  }\href@noop {} {\bibfield  {journal} {\bibinfo  {journal} {Phys. Rev. Lett.}\
  }\textbf {\bibinfo {volume} {99}},\ \bibinfo {pages} {177209} (\bibinfo
  {year} {2007})}\BibitemShut {NoStop}%
\bibitem [{\citenamefont {Wang}\ \emph
  {et~al.}(2011{\natexlab{b}})\citenamefont {Wang}, \citenamefont {Hsieh},
  \citenamefont {Pilon}, \citenamefont {Fu}, \citenamefont {Gardner},
  \citenamefont {Lee},\ and\ \citenamefont {Gedik}}]{Gedik2011}%
  \BibitemOpen
  \bibfield  {author} {\bibinfo {author} {\bibfnamefont {Y.~H.}\ \bibnamefont
  {Wang}}, \bibinfo {author} {\bibfnamefont {D.}~\bibnamefont {Hsieh}},
  \bibinfo {author} {\bibfnamefont {D.}~\bibnamefont {Pilon}}, \bibinfo
  {author} {\bibfnamefont {L.}~\bibnamefont {Fu}}, \bibinfo {author}
  {\bibfnamefont {D.~L.}\ \bibnamefont {Gardner}}, \bibinfo {author}
  {\bibfnamefont {Y.~S.}\ \bibnamefont {Lee}}, \ and\ \bibinfo {author}
  {\bibfnamefont {N.}~\bibnamefont {Gedik}},\ }\href@noop {} {\bibfield
  {journal} {\bibinfo  {journal} {arXiv: 1101.5636}\ } (\bibinfo {year}
  {2011}{\natexlab{b}})}\BibitemShut {NoStop}%
\bibitem [{\citenamefont {Butch}\ \emph {et~al.}(2010)\citenamefont {Butch},
  \citenamefont {Kirshenbaum}, \citenamefont {Syers}, \citenamefont {Sushkov},
  \citenamefont {Jenkins}, \citenamefont {Drew},\ and\ \citenamefont
  {Paglione}}]{Butch_PRB2010}%
  \BibitemOpen
  \bibfield  {author} {\bibinfo {author} {\bibfnamefont {N.~P.}\ \bibnamefont
  {Butch}}, \bibinfo {author} {\bibfnamefont {K.}~\bibnamefont {Kirshenbaum}},
  \bibinfo {author} {\bibfnamefont {P.}~\bibnamefont {Syers}}, \bibinfo
  {author} {\bibfnamefont {A.~B.}\ \bibnamefont {Sushkov}}, \bibinfo {author}
  {\bibfnamefont {G.~S.}\ \bibnamefont {Jenkins}}, \bibinfo {author}
  {\bibfnamefont {H.~D.}\ \bibnamefont {Drew}}, \ and\ \bibinfo {author}
  {\bibfnamefont {J.}~\bibnamefont {Paglione}},\ }\href@noop {} {\bibfield
  {journal} {\bibinfo  {journal} {Phys. Rev. B}\ }\textbf {\bibinfo {volume}
  {81}},\ \bibinfo {pages} {241301} (\bibinfo {year} {2010})}\BibitemShut
  {NoStop}%
\bibitem [{\citenamefont {Kim}\ \emph {et~al.}(2011)\citenamefont {Kim},
  \citenamefont {Cho}, \citenamefont {Butch}, \citenamefont {Syers},
  \citenamefont {Kirshenbaum}, \citenamefont {Adam}, \citenamefont {Paglione},\
  and\ \citenamefont {Fuhrer}}]{Dohun_NatPhys2012}%
  \BibitemOpen
  \bibfield  {author} {\bibinfo {author} {\bibfnamefont {D.}~\bibnamefont
  {Kim}}, \bibinfo {author} {\bibfnamefont {S.}~\bibnamefont {Cho}}, \bibinfo
  {author} {\bibfnamefont {P.~N.}\ \bibnamefont {Butch}}, \bibinfo {author}
  {\bibfnamefont {P.}~\bibnamefont {Syers}}, \bibinfo {author} {\bibfnamefont
  {K.}~\bibnamefont {Kirshenbaum}}, \bibinfo {author} {\bibfnamefont
  {S.}~\bibnamefont {Adam}}, \bibinfo {author} {\bibfnamefont {J.}~\bibnamefont
  {Paglione}}, \ and\ \bibinfo {author} {\bibfnamefont {M.~S.}\ \bibnamefont
  {Fuhrer}},\ }\href@noop {} {\bibfield  {journal} {\bibinfo  {journal} {Nat.
  Phys. doi:10.1038/nphys2287}\ } (\bibinfo {year} {2011})}\BibitemShut
  {NoStop}%
\bibitem [{\citenamefont {Sze}\ and\ \citenamefont {Ng}(2007)}]{Sze_2007}%
  \BibitemOpen
  \bibfield  {author} {\bibinfo {author} {\bibfnamefont {S.~M.}\ \bibnamefont
  {Sze}}\ and\ \bibinfo {author} {\bibfnamefont {K.~K.}\ \bibnamefont {Ng}},\
  }\href@noop {} {\emph {\bibinfo {title} {Physics of Semiconductor Devices}}}\
  (\bibinfo  {publisher} {Wiley},\ \bibinfo {year} {2007})\BibitemShut
  {NoStop}%
\bibitem [{\citenamefont {T\={o}yama}\ \emph {et~al.}(1985)\citenamefont
  {T\={o}yama}, \citenamefont {Takahashi}, \citenamefont {Murakami},\ and\
  \citenamefont {K\={o}riyama}}]{Toyama_APL1985}%
  \BibitemOpen
  \bibfield  {author} {\bibinfo {author} {\bibfnamefont {N.}~\bibnamefont
  {T\={o}yama}}, \bibinfo {author} {\bibfnamefont {T.}~\bibnamefont
  {Takahashi}}, \bibinfo {author} {\bibfnamefont {H.}~\bibnamefont {Murakami}},
  \ and\ \bibinfo {author} {\bibfnamefont {H.}~\bibnamefont {K\={o}riyama}},\
  }\href@noop {} {\bibfield  {journal} {\bibinfo  {journal} {Applied Physics
  Letters}\ }\textbf {\bibinfo {volume} {46}},\ \bibinfo {pages} {557}
  (\bibinfo {year} {1985})}\BibitemShut {NoStop}%
\bibitem [{\citenamefont {T\={o}yama}(1988)}]{Toyama_JAP1988}%
  \BibitemOpen
  \bibfield  {author} {\bibinfo {author} {\bibfnamefont {N.}~\bibnamefont
  {T\={o}yama}},\ }\href@noop {} {\bibfield  {journal} {\bibinfo  {journal}
  {Journal of Applied Physics}\ }\textbf {\bibinfo {volume} {63}},\ \bibinfo
  {pages} {2720} (\bibinfo {year} {1988})}\BibitemShut {NoStop}%
\bibitem [{\citenamefont {Hsieh}\ and\ \citenamefont
  {Gedik}()}]{PersonalHsiehGedik}%
  \BibitemOpen
  \bibfield  {author} {\bibinfo {author} {\bibfnamefont {D.}~\bibnamefont
  {Hsieh}}\ and\ \bibinfo {author} {\bibfnamefont {N.}~\bibnamefont {Gedik}},\
  }\href@noop {} {\bibinfo  {journal} {Personal communication}\ }\BibitemShut
  {NoStop}%
\bibitem [{\citenamefont {Cho}\ \emph {et~al.}(2011)\citenamefont {Cho},
  \citenamefont {Butch}, \citenamefont {Paglione},\ and\ \citenamefont
  {Fuhrer}}]{Sungjae_Nano2011}%
  \BibitemOpen
\bibfield  {journal} {  }\bibfield  {author} {\bibinfo {author} {\bibfnamefont
  {S.}~\bibnamefont {Cho}}, \bibinfo {author} {\bibfnamefont {P.~N.}\
  \bibnamefont {Butch}}, \bibinfo {author} {\bibfnamefont {J.}~\bibnamefont
  {Paglione}}, \ and\ \bibinfo {author} {\bibfnamefont {M.~S.}\ \bibnamefont
  {Fuhrer}},\ }\href@noop {} {\bibfield  {journal} {\bibinfo  {journal} {Nano
  Lett.}\ }\textbf {\bibinfo {volume} {11}},\ \bibinfo {pages} {1925} (\bibinfo
  {year} {2011})}\BibitemShut {NoStop}%
\bibitem [{\citenamefont {Appelbaum}\ \emph {et~al.}(2003)\citenamefont
  {Appelbaum}, \citenamefont {Monsma}, \citenamefont {Rusell}, \citenamefont
  {Narayanamurti},\ and\ \citenamefont {Marcus}}]{SVPD_Appelbaum2003}%
  \BibitemOpen
  \bibfield  {author} {\bibinfo {author} {\bibfnamefont {I.}~\bibnamefont
  {Appelbaum}}, \bibinfo {author} {\bibfnamefont {K.~J.}\ \bibnamefont
  {Monsma}}, \bibinfo {author} {\bibfnamefont {K.~J.}\ \bibnamefont {Rusell}},
  \bibinfo {author} {\bibfnamefont {V.}~\bibnamefont {Narayanamurti}}, \ and\
  \bibinfo {author} {\bibfnamefont {C.~M.}\ \bibnamefont {Marcus}},\
  }\href@noop {} {\bibfield  {journal} {\bibinfo  {journal} {Applied Physics
  Letters}\ }\textbf {\bibinfo {volume} {83}},\ \bibinfo {pages} {3737}
  (\bibinfo {year} {2003})}\BibitemShut {NoStop}%
\end{thebibliography}%

\end{document}